# Symbolic Music Loop Generation with VQ-VAE

**Sangjun Han, Hyeongrae Ihm, Woohyung Lim**

Data Intelligence Lab, LG AI Research
{sj.han, hrim, w.lim}@lgresearch.ai

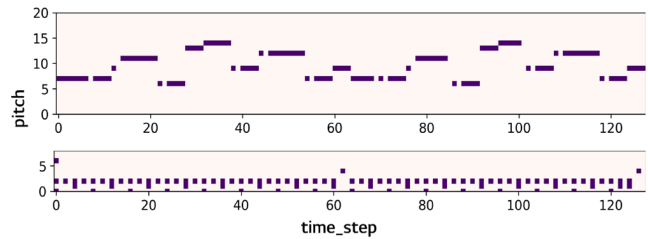

Figure 1. Loop Pianoroll Representation of Bass and Drum

## Abstract

Music is a repetition of patterns and rhythms. It can be composed by repeating a certain number of bars in a structured way. In this paper, the objective is to generate a loop of 8 bars that can be used as a building block of music. Even considering musical diversity, we assume that music patterns familiar to humans can be defined in a finite set. With explicit rules to extract loops from music, we found that discrete representations are sufficient to model symbolic music sequences. Among VAE family, musical properties from VQ-VAE are better observed rather than other models. Further, to emphasize musical structure, we have manipulated discrete latent features to be repetitive so that the properties are more strengthened. Quantitative and qualitative experiments are extensively conducted to verify our assumptions.

## Introduction

With an advance of generative models, many researchers are trying to model long sequences such as language and speech. Music is also a sequence of notes played with multiple instruments, so estimating $p(x)$ for music can be possible. The music generation process can be divided into various levels; audio generation, instrument synthesizer, and composer (Ji et al. 2020). We focus on symbolic music generation since its digital protocol is well defined with standard instruments and note timing information. However, the music sequence is still long and variable that some researchers have transformed the problem into generating 8 or 16 bar phrases (Dong et al. 2018b, Roberts et al. 2018, Gillick et al. 2019).

Music has repetitive structures from motifs to phrases. We address two underlying assumptions; 1) *music can be composed by repeating a certain number of bars in a structured way*. In this respect, the music generation task may be simplified to generate one repetitive pattern which consists of a few bars (we call them loop). Assuming 8 bars as a minimal unit of the loop, the pattern type for each bar can be described as follows; (A - B - A - B - A - B - C - D or A - B - C - D - A - B - C - D or A - B - A - B - A - B - A - B and so on). A common insight from these examples is that "A" should be repeated at regular intervals and referred at the starting point of loops. Using this explicit rule, we created 8 bars loop dataset with additional conditions.

The second assumption is *2) music patterns familiar to humans could be defined in a finite set*. According to Oord's research, they insisted that learning discrete features is sufficient to represent continuous space since many modalities consist of a sequence of symbols (Oord, Vinyals and Kavukcuoglu 2018). Objects in vision, words in language, and phonemes in speech may be candidates of symbols. If we regard each bar as symbol and combine them to be a loop, it is natural to adopt Vector Quantized Variational Autoencoders (VQ-VAE) as our loop generator. The training process of VQ-VAE implies that the frequently referenced patterns are learned automatically from data and stored in a dictionary.

In this work, we propose a loop of 8 bars generation using VQ-VAE. Loop generation is the first step toward the ultimate goal of composing complete music. Since we aim for generating polyphony and multitrack sequences, bass and drum are chosen for our experiments, which are basic components of rhythm and melody. Our contributions can be summarized as follows:

- We propose explicit rules to extract and generate structured a loop of 8 bars which can be used as a building block of music.

- We verify that VQ-VAE is more suitable as a loop generator than other continuous VAE family by representing patterns in discrete space.
- We evaluate the generated samples on new metrics which measure the achievement of loop properties relative to the training set.
- By manipulating discrete latent space to be repetitive, we verify that musical properties that we have intended can be more strengthened.

# Related Work

### Symbolic Music Generation

Symbolic music, known as MIDI, is a communication protocol for digital music instruments. It includes note-on, note-off, velocity, note timing, and duration. To make MIDI available in machine learning, two representation methods are prevalent; event-based representation and time-grid based representation (Ji et al. 2020). The former can represent high time-resolution with a few event vectors and include velocity information. However, a wrong pair between note-on and note-off events can trigger unexpected behavior during music generation. In addition, since events occur sparsely, we cannot benefit from repetitive structures for music generation task. The latter quantizes continuous notes to fixed-length binary vectors. In this work, even if music details may be lost, we choose 16th note quantization because loop structure can be easily observed as Figure 1.

One of the issues of previous studies is how to deal with a polyphony multitrack representation. MusicVAE (Roberts et al. 2018) has applied a shared encoder and three decoders each corresponding to melody, bass, and drum. By adding program-select event to be learnable, Multitrack MusicVAE (Simon et al. 2018) has tried to predict instrument number automatically. In MuseGAN (Dong et al. 2018b), they have fixed instruments with their length and stacked five tracks of time-grid representation. We follow MuseGAN style but modify it to be applicable to both CNN and LSTM models. Various generative models have been reported; sole LSTM decoder (Oore et al. 2018), LSTM encoder and decoders (Roberts et al. 2018), CNN based GAN (Dong et al. 2018b), Seq2Seq Transformer (Huang et al. 2019) and Transformer Autoencoder (Choi et al. 2020). It is known that Transformer (Vaswani et al. 2017) may help to improve generative performance with large datasets (Dosovitaskiy et al., 2021). However, it is difficult to obtain large datasets because MIDI datasets for multitrack are not diverse and well established. In this respect, it is required to have appropriate inductive biases to model music sequences. We have compared CNN and LSTM based Autoencoders which consider temporal dependency to determine which is more appropriate.

Generally, some researchers have prepared their input dataset by sliding window (8 or 16 bars) with a stride of 1 bar. Although they have achieved good performance, this process does not consider relative positions within music, which results in generating ambiguous phrases. To cope with this problem, we focus on generating a loop of 8 bars loops whose repetitive patterns can be clearly recognized.

### VAE and VQ-VAE

VAE is a latent variable model that approximates true posterior, predicting latent **z** on Gaussian distribution and reconstructs it to original data space (Kingma and Welling 2014, Higgins et al. 2017). It can be achieved by maximizing likelihood $p_\theta(x|z)$ parameterized by $\theta$, and minimizing Kullback-Leibler divergence between the approximated posterior $q_\phi(z|x)$ parameterized by $\phi$ and prior $p(z)$. The objective is summarized as

$$\mathbb{E}[\log p_\theta(x|z)] - \beta KL(q_\phi(z|x)||p(z)) \quad (1)$$

where $\beta$ controls channel capacity and independence constraints. The independence, however, does not guarantee the complete separation of musical components that humans can understand. As discussed later, small perturbation on the tangled representation may break our intentions for the loop despite well-prepared dataset.

As opposed to VAE with continuous prior, Oord has proved that a finite set of latent codes is sufficient while it prevents posterior collapse (Oord, Vinyals and Kavukcuoglu 2018). The model, called VQ-VAE, maps a data sequence into a discrete latent space and reconstructs it to the original data space. Similar to equation (1), the objective becomes

$$\mathbb{E}[\log p_\theta(x|z)] - \beta ||q_\phi(z|x) - sg[e]|| \quad (2)$$

where $sg$ denotes a stop gradients operator for the dictionary embedding $e$. Continuous $z$ from encoder $q_\phi(z|x)$ is quantized to nearest embedding $e$. The second term above is responsible for latent $z$ not to diverge far from $e$. For every batch, the dictionary is updated in the sense of centroids of K-means clustering.

A recent study (Razavi et al. 2019) has showed that VQ-VAE with multi-scale discrete features is able to generate high-quality images comparable with the latest GAN models. Later, there were studies related to adopting VQ-VAE for music generation. Jukebox (Dhariwal et al., 2020) has built their VQ model on raw audio domain. Music VQ-CPC (Hadjeres and Crestel 2020) has suggested a similar concept to ours in that they have regarded each discrete feature as a template, but it is difficult to adopt the strategy for sampling negative set because music has recurring patterns. Focusing on symbolic music, we verify our assumptions with reconstruction-based VQ-VAE by evaluating in a quantitative and qualitative manner.

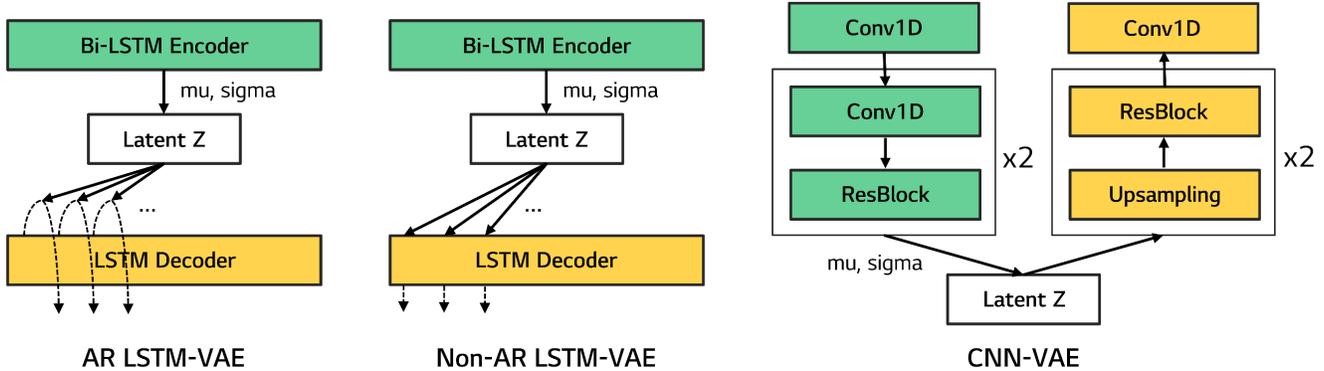

Figure 2. Continuous VAE Family

## Proposed Method

### Data Preprocess

Our objective is to build a generative model of symbolic music (MIDI) with multi-track polyphony style. Lakh MIDI Dataset (LMD) is a collection of MIDI files with various genres and tracks, so it is appropriate to conduct symbolic music experiments (Raffel 2016). In order to standardize the dataset for machine learning, we have to fix the number of notes in a bar, tracks to use, and pitches or components in each track. In this experiment, each note is quantized on the 16th note unit as binary representation so that 16 notes are placed in a bar. To verify the feasibility of multi-track polyphony models, we have extracted two instruments from the dataset; bass guitar (program=32~39) and drum (is_drum=True) which are the core of melody and rhythmic pattern in music. Bass pitches are clipped from C1 to B7 (84 pitches). If several pitches are played at the same time, we have made only the lowest pitch alive for natural bass play. In the case of drum set, 9 components (kick, snare, closed hi-hat, low tom, mid tom, crash, and ride) are regarded as a standard set and the rest of the components are incorporated into the closest standard set or discarded. Formally, our pianoroll representation can be described as follows; $x \in \{0, 1\}^{TB \times P}$ where $T$ is the number of time steps in a bar ($T = 16$), $B$ is the number of bars ($B = 8$), and $P$ is the number of pitches ($P = 93$).

Since most MIDI files in LMD contain full-length songs, it is required to extract 8 bars with a repetitive pattern from them. It is assumed that if the first and fifth bar in the window are highly correlated, the pattern is considered as the loop. While scanning a song with 8 bars striding window, multiple loops can be detected. We summarize and introduce additional conditions to create natural loops as below;

- Hamming distance between the first and fifth bar in a loop should be less than 0.0015. (for loop structure)
- Drum kick and crash should be played on the first note. (to clarify starting point of loop)
- Any pitch of bass should be played on the first note. (to clarify starting point of loop)
- If duplicate bass play occurred, only the lowest pitch is allowed. (for natural bass play)

All the conditions will be referred again in the evaluation section. For 176,581 MIDI files in LMD-full, 150,750 loops are extracted from 114,444 MIDIs by removing non-4/4 signature files. Using pretty_midi (Raffel and Ellis 2014) and pypianoroll (Dong et al. 2018a) in Python library, MIDI processing is conducted.

### Music Continuous VAE

As baseline models for discrete representations, we compared three continuous VAE; autoregressive LSTM-VAE (AR LSTM-VAE), non-autoregressive LSTM-VAE (NonAR LSTM-VAE), and CNN-VAE (Figure 2.). The task for all models is to reconstruct input data, being imposing latent z to Gaussian prior. Our input data representation can be regarded as multi-label class (Read and Perez-Cruz 2014) for each time step (multiple 1s can exist on pitch $P$ axis at a specific time), so cross-entropy loss with softmax is not suitable for Autoencoders. As follows, reconstruction error is denoted as

$$\text{BinaryCrossEntropy}(\text{sigmoid}(logits), \text{targets}) \quad (3)$$

When $\text{sigmoid}(logits) \geq 0.5$, it predicts label as 1, otherwise 0.

AR LSTM-VAE imitates a flat version of MusicVAE (Roberts et al. 2018) which consists of a bidirectional LSTM and autoregressive LSTM decoder. Hidden states $\vec{h}, \overleftarrow{h}$ are obtained from bi-LSTM encoder, which are concatenated and fed into two fully connected layers. Each layer is responsible to produce mean and variance respectively, so that latent features are sampled from them. Layer normalization (Ba, Kiros and Hinton 2016) is applied to its mean outputs. The LSTM decoder, of which hidden states and cells are initialized from sampled latent $z$, autoregressively generates a prediction of next time step with a concatenation $z$ to its output (teacher forcing at training phase and full sampling at inference phase). During training, $\beta$ annealing from 0 to 1 and scheduled sampling (Bengio et al. 2015) with inverse sigmoid rate 20 are adopted.

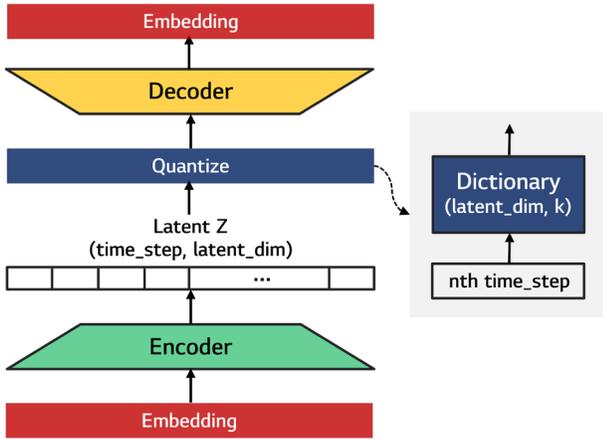

Figure 3. Music VQ-VAE

NonAR LSTM-VAE is little different from the AR version in terms of decoding process. The latent $z$ is replicated for full-time steps and fed into a LSTM decoder so that it produces hidden states for all time steps. After that, they are fed into a fully connected layer to match the target size. Although the NonAR version cannot generate variable-length sequences, we have expected the generated samples to have structural stability by avoiding cumulative error over time.

CNN-VAE can take advantage of computation efficiency rather than RNN model. The encoder consists of three successive blocks including sub-blocks which contain conv_block and residual_block. The conv_blocks are built with (conv1d-batchnorm-leakyrelu(0.2)) and the residual blocks are $x$ + (conv1d-batchnorm-leakyrelu(0.2)-conv1d-batchnorm). The outcomes are flattened and fed into two fully connected layers as we have done for mean and variance in LSTM model. Decoding process is the reverse operation of encoding while upsampling is performed in nearest neighbor mode. Both NonAR LSTM-VAE and CNN-VAE adopt $\beta$ annealing technique from 0 to 1 during training. One can refer to the appendix for more details.

**Music VQ-VAE**

Music VQ-VAE encodes $x = \{x_1, x_2, \cdots, x_{TB}\}, x_i \in \mathbb{R}^P$ into $z = \{z_1, z_2, \cdots, z_t\}, z_i \in \mathbb{R}^L$ where $t$ denotes the number of time step ($t = 32$) in latent space and $L$ denotes latent dimension ($L = 16$). From randomly initialized dictionary $e = \{e_1, e_2, \cdots, e_K\}$, latent $z_i$ is mapped to the nearest embedding $e_k$ where $k = argmin_j \|z_i - e_j\|$. Empirically, we have verified that high compression ratio of latent space, especially for $t$ axis, deteriorates the reconstruction task. For a fair comparison with CNN-VAE, the latent dimensions ($t \times L$) are set to 512 and $K$ also to 512. Our VQ-VAE has similar encoder-decoder structure with CNN-VAE except embedding layer and quantizing layer. The embedding layer, which consists of 2 convolutional layers with kernel size 1, is responsible for reducing pitches to be dense representations and the quantizing layer is for mapping continuous latent features to discrete ones. During training, $e_k$ is updated from the average of near neighbors $z$ set.

| Model | Reconstruction Error |
|---|---|
| AR LSTM-VAE | 3.68 ×1e-3 |
| NonAR LSTM-VAE | 3.37 ×1e-3 |
| CNN-VAE | 3.81 ×1e-3 |
| VQ-VAE | 5.11 ×1e-3 |

Table 1. Reconstruction error. This is calculated as the average of the hamming distance between input and target samples of the validation set.

| $t$ | $L$ | $K$ | Reconstruction Error |
|---|---|---|---|
| 32 | 16 | 256 | 6.05 ×1e-3 |
| 32 | 16 | 512 | 5.11 ×1e-3 |
| 32 | 32 | 512 | 5.06 ×1e-3 |
| 64 | 16 | 256 | 3.89 ×1e-3 |
| 64 | 16 | 512 | 3.33 ×1e-3 |
| 64 | 32 | 512 | 3.02 ×1e-3 |

Table 2. Reconstruction error according to latent space size of VQ-VAE.

To make VQ-VAE be stochastic generative model, autoregressive prior $\prod_{i=0}^{n} p_\theta(z_i | z_{<i})$ over quantized embeddings should be constructed. We have obtained quantized indices $k$ from the VQ-VAE encoder by forward-passing training set to the dictionary and trained new 2-layers LSTM with them. For each time step, softmax output predicts next step index $k$. (Accuracy 82.91 % at teacher forcing mode). $z_0$ is sampled from multinomial distribution $p(z_0)$ of the training set. When sampling unseen data at full sampling mode, temperatures in softmax enable us to control sample diversity.

## Quantitative Evaluation

For quantitative evaluation of generated samples, we have adopted two concepts; *1) model metric* related to evaluating model performance and *2) musical metrics* related to measuring how much our intended properties are involved in generated samples.

| Model | HD | FND | FNB | DB | OS | US | UP | ND |
|---|---|---|---|---|---|---|---|---|
| Training set | 4.43 ×1e-4 | 1.00 | 1.00 | 0.00 | 1.00 | 0.51 | 10.50 | 0.92 |
| AR LSTM-VAE | 4.05 ×1e-3 | 0.99 | 0.44 | 1.51 ×1e-2 | - | - | **10.61** | 0.73 |
| NonAR LSTM-VAE | 2.79 ×1e-3 | **1.00** | 0.41 | 1.89 ×1e-2 | - | - | 10.79 | 0.75 |
| CNN-VAE | 7.33 ×1e-3 | 0.96 | 0.71 | 4.44 ×1e-2 | - | - | 13.81 | 0.77 |
| VQ-VAE (argmax) | **1.44 ×1e-3** | 1.00 | **0.98** | **1.10 ×1e-3** | 0.28 | 4.08 ×1e-3 | 8.77 | 0.94 |
| VQ-VAE (temp 1) | 4.25 ×1e-3 | **1.00** | **0.98** | 2.21 ×1e-3 | 7.97 ×1e-2 | 0.50 | 9.94 | **0.91** |
| VQ-VAE (temp 1.5) | 1.23 ×1e-2 | **1.00** | **0.98** | 2.56 ×1e-3 | 2.45 ×1e-2 | 0.90 | 11.58 | 0.89 |
| VQ-VAE (temp 2) | 1.82 ×1e-2 | **1.00** | 0.97 | 3.09 ×1e-3 | **2.89 ×1e-3** | **0.99** | 14.05 | 0.88 |

Table 1. Musical Metric Results

| Model | HD | FND | FNB | DB | OS | US | UP | ND |
|---|---|---|---|---|---|---|---|---|
| VQ-VAE (argmax) | 1.56 ×1e-3 | 1.00 | 0.98 | 1.01 ×1e-3 | 0.25 | 4.11 ×1e-3 | 8.69 | 0.95 |
| VQ-VAE (temp 1) | 3.58 ×1e-3 | 1.00 | 0.98 | 2.18 ×1e-3 | 7.12 ×1e-2 | 0.62 | 10.02 | 0.91 |
| VQ-VAE (temp 1.5) | 6.89 ×1e-3 | 1.00 | 0.98 | 2.61 ×1e-3 | 2.25 ×1e-2 | 0.92 | 11.44 | 0.90 |
| VQ-VAE (temp 2) | 9.99 ×1e-3 | 1.00 | 0.97 | 3.14 ×1e-3 | 2.95 ×1e-3 | 0.99 | 13.72 | 0.88 |

Table 2. Musical Metrics by Manipulating Latent Space

**Model Metric**

**Reconstruction Error:** For both VAE and VQ-VAE, reconstruction error is part of the objective function that indicates how well the model decodes its latent codes to original space. However, perfect satisfaction of the objective does not guarantee generating high-quality samples. VAE has produced many noisy samples even after achieving the minimal reconstruction error when $\beta \approx 0$. It indicates that the posterior $q_\phi(z|x)$ failed to fit Gaussian prior. We have extensively searched and selected the best condition ($\beta = 1$) in terms of musical metrics introduced later (without any mention, $\beta$ is scheduled to be 1). As referred in Oord's research (Oord, Vinyals and Kavukcuoglu 2018), $\beta$ in VQ-VAE objectives did not affect much to reconstruction task ($\beta = 0.25$).

**Musical Metrics**

**Loop Properties:** The four metrics newly suggested are related to our data preprocessing. Since the generative models are trained with the loop dataset that we have created (we have mentioned that the loop phrase should have a repetitive structure and a clear starting point), they are expected to include four features in their outcomes.

- *ham_dist (HD)*: the average of the hamming distance between the first and fifth bar for all samples. (for loop structure)
- *first_note_drum (FND)*: the ratio of samples including kick and crash appearing on the first note. (to clarify starting point of loop)
- *first_note_bass (FNB)*: the ratio of samples including bass play appearing on the first note. (to clarify starting point of loop)
- *dup_bass (DB)*: the ratio of duplicate bass play for each time step. (for natural bass play)

**Creativeness (for only VQ-VAE):** Merely memorizing and fetching samples from the training set is not a creative process. Especially for VQ-VAE, restricting latent space on a finite number may deteriorate sample diversity. Here, we have evaluated creativeness by comparing generated samples on latent space to avoid large matrix computation. Since latent features from VAE family are continuous, we excluded the experiments related to these metrics.

- *overlap_sample (OS)*: the ratio of generated samples overlapped with the training set.
- *unique_sample (US)*: the ratio of uniqueness among generated samples.

**Musical Style:** Musical style can be inferred from the number of used pitches and rhythm patterns. The two metrics below are inspired by this research (Choi et al. 2020).

- *unique_pitch (UP)*: the average number of used pitches per bar.
- *note_density (ND)*: the average number of notes per bar.

## Experiments and Results

### Experiment Details

To evaluate musical properties, our experimental models generated the same number of samples compared to the training set (training set: 120,600, validation set: 15,075, test set: 15,075). Continuous VAE decodes samples from Gaussian prior, and VQ-VAE does from the LSTM learned prior. We have forced LSTM models to have latent size 128 and CNN models to have 512-dimensions. We have trained them with 2,000 epochs and batch size of 2,048. During training, AdamW optimizer (Loshchilov and Hutter 2019) is adopted with cosine annealing from 1e-3 to 5e-6. All models are initialized with He initializer (He et al. 2015).

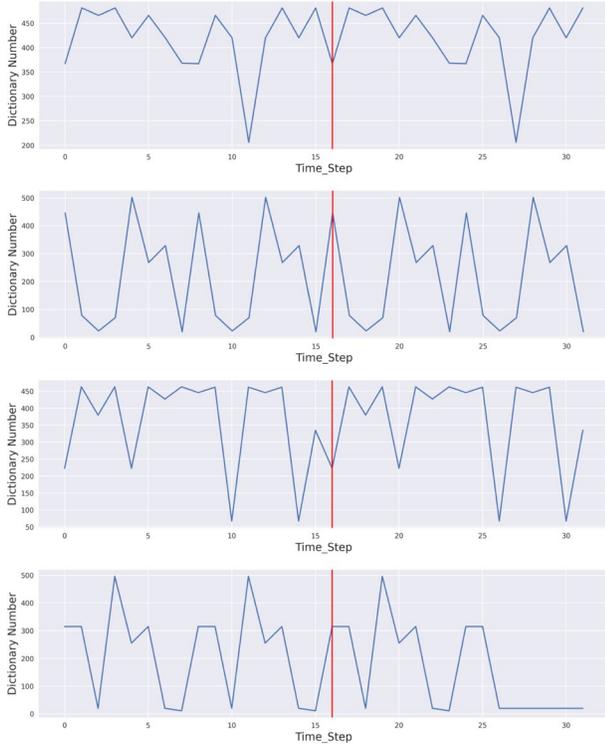

Figure 4. Loop patterns in discrete latent space. The x axis denotes compressed time dimension and the y axis denotes quantized index $k$ from dictionary.

**Model Metric Results**

Due to the finite latent codes, VQ-VAE is a little worse than the continuous VAE family in terms of reconstruction task (Table 1.). It depends heavily on the compression ratio of latent space and the number of classes in the dictionary (Table 2.). Despite of same latent space size, it seems that reducing time step dimension ($t$) makes worse the task in VQ-VAE. Although the high compression ratio deteriorates reconstruction quality, it makes modeling prior easier by reducing data space to train. We have fixed $t = 32, L = 16, K = 512$ and conducted later experiments.

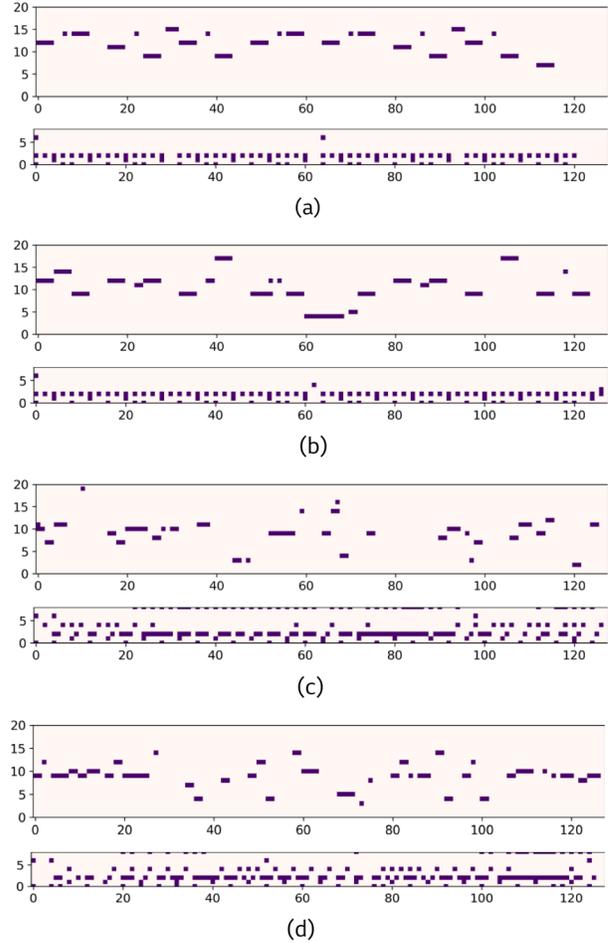

Figure 5. Investigation of frequent samples on data space. Samples (a), (b) are first and second most frequently appeared and (c), (d) are first and second least frequently appeared for each latent time steps.

**Musical Metric Results**

This section is for interpreting Table 3. Except *for OS* and *US*, musical metrics from the training set are desirable values for our generated set (top row). The two exceptions are indicators to check sample uniqueness of intra or inter group, so it is preferred to have smaller values. Comprehensively, VQ-VAE with argmax sampling from prior shows good performance in terms of *HD*, *FND*, *FNB*, and *DB*. However, only 0.4 % of samples from generated samples are internally unique and 28 % of samples are overlapped with the training set. We can improve it by relaxing argmax to probability sampling with temperature control. VQ-VQE (temp 1) is a little worse than NonAR LSTM-VAE in terms of *HD* and *UP* but shows much better in *FNB*, *DB,* and *ND*. As for *OS*, almost 8 % of samples are overlapped. *US* results have showed that 50 % of samples are unique and it is comparable with ones in the training set. The reason *US* 0.51 from the training set is that a song can contain several repetitive loops. By applying higher temperatures, the uniqueness can be strengthened while sacrificing other metrics.

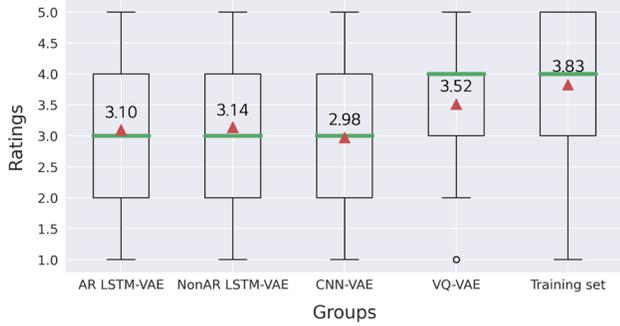

Figure 6. Ratings boxplot from listening test. Green line and red triangle indicate median and mean (above indicating mean value).

It is noticed that *FNB* and *DB* in continuous VAE has decreased dramatically. Although all training data are processed to play bass at the first note, it seems that the bass structure collapses in latent space. In contrast, all models maintain drum structure well (*FND*). About *HD*, NonAR LSTM-VAE has received the lowest *HD* value except VQ-VAE (argmax). The model with bi-directional and repetitive features keeps consistency well between the first and fifth bar. In terms of *ND*, it has showed that continuous VAE tends to play fewer notes than VQ-VAE.

## Manipulating Latent Space of VQ-VAE

Unlike other models, VQ-VAE keeps temporal information in 2D latent space. As Figure 4., the loop patterns in discrete latent space can be clearly observed. Taking advantage of these properties, we can improve the musical metrics. We copied latent features at latent time step 0, 1 to 16, 17 so that consistency between the first and fifth bar can be strengthened. Except for VQ-VAE (argmax), *HD* decreased without sacrificing other metrics (Table 4.). This process is only applicable to VQ-VAE.

## Investigation of Frequent Pattern

Dictionary embeddings behave like centroids in K-means clustering, so it is worth investigating their representation in terms of frequency of appearance. For each latent time step, we have picked quantized indices that appear most and least frequently and represent them on data space (Figure 5.). As Figure 5. (a) and (b), it is observed that frequently referenced indices show repetitive and typical patterns for 4/4 signature. Nevertheless, the samples sound unnatural since the indices are sampled depending on frequency without considering any temporal dependency. For samples (c) and (d), it is difficult to find musical sources to use for music.

## Listening Test

We have tried to capture music properties through model and musical metrics, but it is difficult to define human perception in a quantitative manner. To overcome the limitation, we have conducted a listening test for 20 people. We have selected the training set as a baseline and compared loop samples from four generative models (AR LSTM-VAE, NonAR LSTM-VAE, CNN-VAE, VQ-VAE with temp 1.2). Participants are asked to listen 10 samples for each group (total 50 samples from 5 groups) and evaluate how much the sample sounds like music on a Likert scale. In advance, we have also provided 3 samples them to be familiar with loop patterns.

As Figure 6., the group of training set show the highest ratings with an average rating 3.83. Among generative models, VQ-VAE got the highest average rating 3.52. It seems that generated patterns from discrete representations are repetitive and structural, so the participants felt them closer to the real music. Additionally, we have carried out Kruskal-Wallis H test which is non-parametric one-way ANOVA. The test showed a statistically significant difference among test groups with $H = 75.19, p < 0.001$. We can conclude that VQ-VAE is relatively suitable model for symbolic music generation.

## Discussions

As far as we know, this is the first time to extract explicit patterns from music. Since the process enables us to prepare dataset which is clearly identified with repetitive patterns and their boundary of loop phrase, we can expect generated samples to show the loop structure.

Since proposed models are designed to take polyphony and multi-track symbolic music, they can be easily applied to any number of instruments. From quantitative and qualitative analysis, we have verified that VQ-VAE is suitable to model and generate structural music sequences. Representing music in discrete latent space can benefit from reusing familiar patterns and manipulating time dependency. Other continuous VAE models had difficulty reproducing some musical properties despite of good reconstruction. Our experiments are extensively verified through existing and new metrics appropriate to the music loop. Concerning the uniqueness of generated samples from VQ-VAE, we can control sample diversity by adjusting temperatures in autoregressive prior. We have verified that VQ-VAE with probability sampling can generate human-friendly patterns through the listening test.

In the future, it is more desirable to extract loop structure automatically in the training process of neural networks beyond attention. Also, multi-scale features (Razavi et al. 2019) will be beneficial to generate realistic music.

# Acknowledgments

This research was supported by LG AI Research.

## Appendix A  Implementation Details

### Network Architectures

We introduce network architectures and their hyperparemters used for the experiments. (AR LSTM-VAE, NonAR LSTM-VAE, CNN-VAE and VQ-VAE). For AR LSTM-VAE (Table (a)), LSTM decoder accepts latent $z$ from encoder as initializer of hidden and cell states and autoregressively generates next times step after concatenating the latent $z$ to inputs. CNN based models are based on conv_block (conv1d-batchnorm-leakyrelu(0.2)) and residual_block ($x$ + (conv1d-batchnorm-leakyrelu(0.2)-conv1d-batchnorm)). In tables, conv_block is denoted in the form of conv(channel, kernel_size, padding, stride) and residual_block denoted as resblock(channel, kernel_size, padding, stride).

### Training Details

Our experiment results do not depend on random initialization of neural networks so much. We strictly separated MIDI dataset into train, validation and test set without any shuffling since loops can be duplicated in a MIDI. Each model is trained with NVIDIA GeForce RTX 3090 less than 8 hours. Our implementation works on Pytorch (==1.9.0), Numpy (==1.20.1), pretty_midi (==0.2.9), pyFluidSynth (==1.3.0), pypianoroll (==1.0.4) and scipy (==1.6.2).

### How to Play Pianoroll Representation

As referred in Ji et al. 2020, time-grid based representation cannot distinguish between long notes and repeated notes. If notes are repeated successively, we regarded them as continuous one. Note velocity is set to 80 for bass and 100 for drum. Wav files for our examples are generated through scipy.io.wavfile.write(pretty_midi.fluidsynth()).

**Input:** $x \in \{0, 1\}^{128 \times 93}$

| Model | Operations | Output Shape |
|---|---|---|
| Encoder | $x, (h, c) = \text{Bi-LSTM}(x)$ | $h$ shape: (1, 256) |
| | $\mu = \text{LayerNorm}(\text{Linear}(h))$ | $\mu$ shape: (128) |
| | $\sigma = \text{Linear}(h)$ | $\sigma$ shape: (128) |
| | $z = \text{Reparam}(\mu, \sigma)$ | $z$ shape: (128) |
| Decoder | At time = 0, $h, c = z$ $x_0, (h, c) = \text{LSTM}(0, (h, c))$ At time = 1...127, $x_t = \text{concat}(x_t, z)$ $x_t, (h, c) = \text{LSTM}(x_t, (h, c))$ | $x_t$ shape: (128) |
| | $x_t = \text{Sigmoid}(\text{Linear}(x_t))$ | $x_t$ shape: (1, 93) |

**Output:** $x \in \{0, 1\}^{128 \times 93}$

(a) AR LSTM-VAE

**Input:** $x \in \{0, 1\}^{128 \times 93}$

| Model | Operations | Output Shape |
|---|---|---|
| Encoder | $x, (h, c) = \text{Bi-LSTM}(x)$ | $h$ shape: (1, 256) |
| | $\mu = \text{LayerNorm}(\text{Linear}(h))$ | $\mu$ shape: (128) |
| | $\sigma = \text{Linear}(h)$ | $\sigma$ shape: (128) |
| | $z = \text{Reparam}(\mu, \sigma)$ | $z$ shape: (128) |
| Decoder | $z = \text{repeat}(z)$ | $z$ shape: (128, 128) |
| | $x = \text{LSTM}(z)$ | $x$ shape: (128, 256) |
| | $x_t = \text{Sigmoid}(\text{Linear}(x_t))$ | $x_t$ shape: (1, 93) |

**Output:** $x \in \{0, 1\}^{128 \times 93}$

(b) NonAR LSTM-VAE

**Input:** $x \in \{0, 1\}^{128 \times 93}$

| Model | Operations | Output Shape |
|---|---|---|
| Encoder | $x = \text{conv}(32, 3, 1, 1)(x)$ | $x$ shape: (128, 32) |
| | $x = \text{conv}(32, 3, 1, 2)(x)$ $x = \text{resblock}(32, 3, 1, 1)(x)$ | $x$ shape: (64, 32) |
| | $x = \text{conv}(32, 3, 1, 2)(x)$ $x = \text{resblock}(32, 3, 1, 1)(x)$ | $x$ shape: (32, 32) |
| | $\mu = \text{LayerNorm}(\text{Linear}(x))$ | $\mu$ shape: (512) |
| | $\sigma = \text{Linear}(x)$ | $\sigma$ shape: (512) |
| | $z = \text{Reparam}(\mu, \sigma)$ | $z$ shape: (512) |
| Decoder | $x = \text{Linear}(z)$ | $x$ shape: (512) |
| | $x = \text{upsample}(2)(x)$ $x = \text{resblock}(32, 3, 1, 1)(x)$ | $x$ shape: (32, 32) |
| | $x = \text{upsample}(2)(x)$ $x = \text{resblock}(32, 3, 1, 1)(x)$ | $x$ shape: (64, 32) |
| | $x = \text{upsample}(2)(x)$ $x = \text{conv}(93, 7, 3, 1)(x)$ $x = \text{Sigmoid}(x)$ | $x$ shape: (128, 93) |

**Output:** $x \in \{0, 1\}^{128 \times 93}$

(c) CNN-VAE

| Input: $x \in \{0,1\}^{128 \times 93}$ | | |
|---|---|---|
| Model | Operations | Output Shape |
| Encoder | $x = \text{conv}(64, 1, 0, 1)(x)$<br>$x = \text{conv}(32, 1, 0, 1)(x)$ | $x$ shape: (128, 32) |
| | $x = \text{conv}(32, 3, 1, 2)(x)$<br>$x = \text{resblock}(32, 3, 1, 1)(x)$ | $x$ shape: (64, 32) |
| | $x = \text{conv}(32, 3, 1, 2)(x)$<br>$x = \text{resblock}(32, 3, 1, 1)(x)$ | $x$ shape: (32, 32) |
| | $x = \text{conv}(32, 3, 1, 2)(x)$<br>$x = \text{resblock}(32, 3, 1, 1)(x)$ | $x$ shape: (32, 16) |
| Quantize | $x = \text{quantize}(x)$ | $x$ shape: (32, 16) |
| Decoder | $x = \text{conv}(32, 3, 1, 1)(z)$ | $x$ shape: (32, 32) |
| | $x = \text{upsample}(2)(x)$<br>$x = \text{resblock}(32, 3, 1, 1)(x)$ | $x$ shape: (64, 32) |
| | $x = \text{upsample}(2)(x)$<br>$x = \text{resblock}(32, 3, 1, 1)(x)$ | $x$ shape: (128, 32) |
| | $x = \text{conv}(64, 1, 0, 1)(x)$<br>$x = \text{conv}(93, 3, 1, 1)(x)$ | $x$ shape: (128, 93) |
| | $x = \text{Sigmoid}(x)$ | $x$ shape: (128, 93) |
| Output: $x \in \{0,1\}^{128 \times 93}$ | | |

(d) VQ-VAE